\documentclass[12pt]{article}
\usepackage[a4paper, total={8.5in, 11in}, margin=1in]{geometry}
\usepackage[utf8]{inputenc}
\pdfoutput=1
\usepackage{subcaption}
\usepackage{graphicx}
\usepackage{color}
\usepackage{bm}
\usepackage{amsmath}
\usepackage{mathtools}
\usepackage[normalem]{ulem}
\usepackage[breaklinks=true]{hyperref}

\usepackage[style=nature]{biblatex}
\bibliography{references.bib}

\title{Quantum Clock Synchronization for Future NASA Deep Space Quantum Links and Fundamental Science}

\author{
  James Troupe\textsuperscript{1}\\
  \and
  Stav Haldar\textsuperscript{2}\\
  \and
  Ivan Agullo\textsuperscript{2}\\
  \and
  \and
  Paul Kwiat\textsuperscript{3}\\
}

\date{}
\begin{document}

\maketitle
\thispagestyle{empty}

\begin{center}

Affiliation: \textsuperscript{1}Xairos Systems, Inc., \textsuperscript{2}Louisiana State University,
\textsuperscript{3}University of Illinois at Urbana-Champaign

\hfill \break
Endorser: Dr. Jon Hamkins (Jet Propulsion Laboratory, California Institute of Technology)

\hfill \break
We acknowledge Dr. Daphna Enzer for helpful comments on earlier versions of this paper.

\hfill \break
Primary Point of Contact: James Troupe, (737) 232-0450, james@xairos.com

\hfill \break
Topical white paper submitted for the Decadal Survey on Biological and Physical Sciences Research in Space 2023-2032. Primary contact: james@xairos.com
\end{center}





\newpage

\section*{\small{Abstract:}}
\setcounter{page}{1}

The ability to measure, hold and distribute time with high precision and accuracy is a foundational capability for scientific exploration. Beyond fundamental science, time synchronisation is an indispensable feature of public and private communication, navigation and ranging, and distributed sensing, amongst others technological applications. 

We propose the implementation of a quantum network of satellite- and ground-based clocks with the ability to implement Quantum Clock Synchronization to picosecond accuracy. Implementation of the proposed QCS network offers a double advantage: (1) a more accurate, robust, and secure time synchronization network for classical applications than currently possible, and (2) a resource to fulfill the much more stringent synchronization requirements of future quantum communication networks.

\section{Quantum Clock Synchronization}


In this paper we are primarily interested in the utilization of quantum entanglement as a resource for clock synchronization, Quantum Clock Cynchronization (QCS). 
While most of the original QCS proposals were based on entangling qubit states
of distant optical atomic clocks \cite{Jon_QCS_2018,QClock_network_sat}, we focus here on using entangled photons to connect atomic clocks. Such photon sources are available with current technology and will be required for future quantum networks. 
%
%
Importantly, some entangled photon QCS protocols require only a very few detected photon pairs (about 20 pairs) \cite{AntiaQCS, Lee2019}, making them suited for deep-space links or other high loss links. Optical communication techniques using single-photon detection (non-coherent/quantum measurement) have an advantage in power efficiency (bits per photon) over coherent optical communication techniques \cite{Dolinar2011}, which are generally used in high precision, classical optics-based clock synchronization, e.g., classical Optical Two-Way Time and Frequency Transfer (O-TWTFT). 

\label{2.1}

We provide now a brief description of a concrete realization of a quantum clock synchronization protocol that utilizes the quantum properties of light and has the advantage of being feasible with current technology \cite{Lee2019}. The signals used in classical protocols  are replaced with pairs of individual photons created via spontaneous down-conversion (SPDC) by pumping a nonlinear optical crystal. Due to conservation of energy, the time of birth of the photons are very highly correlated with each other, typically on the order of 10--100 femtoseconds \cite{spdc_time_corr1970, Shih2004}. The SPDC process is also quantum mechanically random with photon pair production following a Poisson distribution. The photon pair production time is itself used as the (now truly) random ``code'' shared between Alice and Bob. Alice and Bob each have an SPDC source and locally detect one photon of each pair produced, recording a timestamp of each detection. The other photon  is  transmitted to the other party and its detection is timestamped. 
The clock offset is then estimated by the difference of the two cross-correlation peaks, as with the classical O-TWTFT scheme. In the case of a relative frequency difference between the two clocks, this method can be extended to also estimate the frequency difference by estimating the change in the clock offsets as a function of the local time of one of the clocks. 

Since this is a type of TWTFT scheme, it assumes ``reciprocity'', namely that the time of flight of photons from Alice to Bob and vice-versa are the same. In classical protocols using RF optical signals, and for the case where Alice and Bob are  a ground station and a satellite in LEO, respectively, the fluctuations in signal multi-path propagation can create channel non-reciprocity of the order a nanosecond. In contrast, with optical photons this figure goes down to the order of a few femtoseconds \cite{Belmonte2017, Taylor2020}. 

The precision of this QCS time distribution method has been demonstrated to be on the order of 10's of picoseconds with low-cost avalanche detectors and time tagging hardware \cite{AntiaQCS, Lee2019}. Single-digit picosecond precision should be readily achievable with faster time tagging and detectors with lower jitter. Simulations show that an accurate clock offset can be produced with fewer than 100 detected down-converted photon pairs with realistic detector dark count rates of about 1 kHz. We note that although distribution protocols using stabilized optical frequency combs are capable of achieving a higher precision of femto- to sub-femtosecond levels, they have higher resource requirements and also generally need pre-synchronisation at the picosecond level \cite{Newbury2016_freq_combs}. It is thus desirable to have a low-cost, low-SWaP (Size Weight and Power), robust, and secure protocol that provides picosecond precision and is stable over periods in which the finer synchronisation is done. QCS using entangled photon pairs is thus an appropriate candidate for this role.

If the SPDC source is configured to also produce polarization entanglement between the photon pairs, then violation of Bell's inequality can be used to directly authenticate the detected photons, providing an extra layer of security similar to the one used in quantum key distribution protocols. Characterizing the achievable secure accuracy of QCS based time distribution in practical realizations is the focus of ongoing research.

\section{Satellite Based QCS and Quantum Networks}\label{3}

Implementation of quantum protocols over large distances has seen tremendous development in the recent years. Quantum state transfer protocols such as teleportation and key distribution have been carried out over continental distances of more than more than 200 km, with satellite-linked nodes separated by more than 1000 km. A key role in developing these implementations has been played by the transition to hybrid space-terrestrial quantum communication network architectures combining satellites and ground stations equipped with optical telescopes with metropolitan-scale fiber optic networks \cite{LSU_satsim}. To contrast with the limits of fiber optic-based quantum communication which is hindered by the exponential losses, we note that the quantum repeater-less fiber optic based secret key rate bound is surpassed beyond 215 km for a satellite at altitude of 530 km \cite{Pan_exp_crypto, makan_review}.
Unlike the classical information encoded in classical optical signals, the quantum information encoded via quantum communication protocols 
cannot be amplified due to fundamental limits on copying quantum information. This places fundamental limits on directly transferring quantum information through lossy channels. Large number of high fidelity quantum repeaters and/or quantum memories could improve the situation to some extent, but their current performance levels are below those needed for mature applications \cite{makan_review}, and furthermore, it would very likely be impractical to place such devices in difficult terrain, e.g. under the ocean. In addition, low altitude fast-moving satellites can provide continental scale coverage areas between ground stations within a small fraction of the day (or very large, more persistent coverage areas using higher altitude satellites with larger telescopes). Thus, the use of inter-satellite and satellite-ground links has great potential to increase the performance and utility of quantum networks over global distances.

Recent proof-of-principle experiments through the Micius satellite have indeed shown the effectiveness of a satellite-based quantum communication channel for large scale quantum networks \cite{Pan_exp_crypto,Pan_exp_teleport,Pan_satellite_QCS}. The clock synchronization jitter in this experiment was in the range of 0.1--1 ns (avg.\ error 0.7 ns). Note that this clock synchronization was performed with a one-way protocol using classical intensity laser pulses from the ground station. Since teleportation of quantum states is a fundamental task of any quantum network, the timing requirements for teleportation set fundamental requirements for the precision of clock synchronization for realising a quantum network at large scales. Picosecond-level synchronization is likely needed for quantum networks to function efficiently \cite{GE_2021}, which is currently not achievable over very long distances using only local timekeeping as GPS only allows synchronization to an accuracy of 20--40 nanoseconds.

\subsection{Future Global Navigation Satellite Systems (GNSS)}

The convergence of low-cost nano- and micro-satellites, low-cost launch systems, maturation of low-SWaP atomic clocks \cite{NIST2020,GPSWorld2011,DARPA2019,DSAC2021}, and the development of compact, efficient entangled photon sources \cite{Bedington2016} have made high precision time distribution (on the order of 1 to 10 picoseconds) on a global scale both feasible and commercially viable. By combining a constellation of small nano- or mico-satellites with on-board low SWaP atomic clocks, compact bright entangled photon sources, and optical links, the quantum clock synchronisation protocol can be used to synchronize the clocks across the constellation and reference them to more capable (e.g., higher precision and stability) ground-based reference clocks. Such a time distribution network would allow a more precise and secure time reference than is currently available with GPS or other global navigation satellite system. This common time resource could provide new capabilities in telecommunication (e.g., higher bandwidth and lower latency), distributed computing, and coherent distributed sensing (e.g., improving sensitivity of very long baseline interferometry). Additionally, implementation of many quantum networking protocols require precision time-bin encoding and measurement coincidence windows. Thus, beyond enabling measurement of weak gravity effects on quantum systems at large scales, such a precision time network would significantly advance a range of technologies and lay the foundation for global scale entanglement distribution and fully quantum networks \cite{QClock_network_sat, Pan_satellite_QCS,LSU_satsim}.

\subsection{Quantum Internet}
The Quantum internet describes an extended network of quantum devices/nodes linked via quantum communication channels which transfer quantum states and entanglement across the network. Its wide range of applications include distributed quantum computing, communication security, distributed sensing, optical VLBI, etc. \cite{Kimble_quantum_internet, LSU_satsim}. A QCS network is ideally suited to fulfill the sub-nanosecond time synchronization requirements of large-scale quantum networks. Just like the Network Time Protocol (NTP) formed the basis for the current internet, a QCS network is a promising near-term foundational implementation of a global scale entanglement-based quantum network \cite{NTP_Mills}. NTP is an essential part of the Internet’s function and is crucial for several classical security protocols such as digital signatures and transport layer security \cite{NTP_security, NTP_Attacks}. The Quantum Internet promises to provide supra-classical security using quantum entanglement. This security will similarly be reliant on the security of the time distribution protocol it uses. A key concept behind the success of the NTP is its hierarchical structure which makes it robust against node failure \cite{NTP_Mills, wiki:Network_Time_Protocol}. QCS can easily adapt such a structure by distributing satellite resources commensurate with the strata to which a clock belongs. E.g., Lower strata client clocks can be serviced by larger constellations of low cost (low stability) LEO orbits whereas costlier and more precise server clocks can be serviced using smaller MEO or GEO constellations. The inter-satellite links create a master clock in the sky with different strata of precision, while satellite-ground and ground-ground links distribute this standard across the globe. 

\subsection{Relativistic Effects in High Precision Satellite Based QCS}
A QCS network can achieve precision significantly greater than the GPS, and hence must account for even tinier relativistic corrections than those considered for classical PNT systems. The subset of these corrections which must be accounted for depends on the time window needed for synchronization. Further, the QCS network is also an ultra-precise measurement tool for higher order relativistic effects which accumulate to measurable differences only over several sync events. 
We now list some of the relevant relativistic effects (for details see \cite{Zhu_GPS_relativity, Ashby_GPS_relativity}). At the threshold of current PNT precision levels lie effects like Shapiro delay which cause travel time delays about 100 picoseconds (or few centimeters) during satellite to ground transit. This is caused by slowing down of EM waves near a gravitational source (like Earth). Next in terms of size is a similar but higher-order time delay effect caused by space-time curvature near a gravitating body. This is of the order of few picoseconds (or few millimeters). Perturbative effects such as those due to orbit corrections, Earth’s non-spherical mass distribution and tidal forces from solar system objects cause time delays ranging from tens to tenths of a picosecond over the course of a day. All these effects are in range of measurement through a near-term implementation of the satellite-based QCS network proposed here. Looking forward, an advanced implementation of the QCS network, also including DSQLs could measure femtosecond-level time delays. This would bring into grasp the measurement of yet elusive gravitomagnetic effects such as the geodetic and Lense-Thirring effects. Since Earth-based atomic clocks have already achieved attosecond precision and sub-femtosecond clock sync capabilities, these effects might become relevant in any attempt to define a global Earth fixed coordinate time at that precision \cite{SAGE, QClock_network_sat}.

\subsection{Fundamental Science and Deep Space Quantum Links}

The way quantum mechanics and  gravitation interact  is a prominent open problem in modern physics. Efforts on this direction, such as quantum field theory in curved space time, need inputs from experiments involving interaction of quantum systems with gravity. 
The QCS network proposed here, whose building blocks are entanglement-based quantum communication channels near gravitational sources, offers an arena germane for such experiments \cite{Qoptics_space_exp}. 
Deep Space Quantum Links, envisaged with an aim to measure these effects, also need an independent and precise time standard which can be provided by the implementation of QCS. 


We now mention some recent proposals to explore the interaction of quantum systems with gravity. Primary amongst these are matter and optical interferometry experiments conducted in space. 
At the heart of all sensitive quantum interference experiments lies the ability to measure and hold time at high precision, hence making the QCS protocol a useful tool. 

The matter 
interferometry proposals includes the study of gravitational phase shifts on a beam of neutrons \cite{neutron_gravAbele2012}, interference experiments using ultra cold matter (Bose-Einstein Condensates --- highly quantum-correlated states) \cite{BEC_space_interference_Lachmann2021} and the Space Atomic Gravity Explorer (SAGE) \cite{SAGE}. The last amongst these is an ambitious proposal to explore effects of gravity on quantum correlations (via Bell tests) and test the limits of existing theories of classical gravity (equivalence principle). 
One of the aims that lies at the heart of SAGE's success is ``Defining an ultra-precise frame of reference for Earth and Space, comparing terrestrial clocks, using clocks and links between satellites for optical VLBI (very long baseline interferometry) in Space" \cite{SAGE}.
Other promising proposals include experiments involving two quantum massive systems (masses greater than $10^{-12}$ kg) gravitationally interacting   and each passing through a matter-wave Mach-Zehnder interferometer \cite{Vedral_QG_test_PRL} (along the lines of the COW experiment \cite{COW}, also see \cite{SBose_QG_test_PRL} for a similar proposal). The role of QCS here is evident, since two different interference experiments cannot co-determine quantum correlations without establishing a common time. 
Promising investigations involving optical interferometry include Mach-Zender interference experiments with one beam-splitter placed on Earth and the other on a satellite \cite{Qoptics_space_exp} and a more sensitive experiment using a ground-based Sagnac type interferometer to measure frame dragging effects on the Hong-Ou-Mandel effect. This offers a clear demarcation between classical vs quantum and Newtonian vs general relativistic regimes \cite{Anthony_HOM_grav}. Other proposals include a study of the effects of weak gravity on Bell tests \cite{Bell_test_weak_grav} and a high precision test of the Einstein's equivalence principle using quantum optical systems \cite{Einstein_eq_Qopt}. Other space based attempts not reliant on interferometry, include the MICROSCOPE mission (a micro-satellite based experiment which measures discrepancies in free fall accelerations) \cite{Microscope_Touboul2017} and the Micius satellite which tested the theoretically expected persistence of entanglement in non-inertial frames with a time precision of tens of picoseconds using Bell tests \cite{Xu_sat_grav_decoherence}.

\section{Conclusion}

It is becoming clear that quantum communication, quantum sensing, and quantum computing are rapidly developing into practical technologies that will impact society over the next few decades. The first applications of these emerging technologies will almost certainly be to enable new scientific instruments which will open up new avenues of discovery and to new methods for processing and understanding data. One of the crucial first steps towards implementing these new technologies will be to establish high accuracy (sub-nanosecond) time distribution infrastructure across the Earth and beyond. 

Because distribution of quantum entanglement is both the fundamental capability necessary for implementing quantum technologies and also enables efficient methods for precision time distribution, we argue that the first steps towards integrating quantum networks into NASA's communication infrastructure should emphasize utilizing entangled photon-based quantum clock synchronization. The technology to implement QCS is mature enough now to deploy on demonstration missions, and subsequently inform the development of QCS requirements and planning for incorporation into future systems. These early QCS systems will connect and leverage the last few decades of research and development that has produced optical atomic clocks with exquisite performance, and which may be used to redefine the second in the next decade. Once a nascent QCS based sub-nanosecond level time distribution infrastructure connecting high stability atomic clocks is accessible to new scientific instruments and space missions, it will have the potential to accelerate scientific discovery. In addition, a highly accurate, secure global time distribution network would enable many terrestrial applications for scientific exploration, commercial utility, and national defense. 

\printbibliography 

@article{makan_review,
Author = {Jasminder S. Sidhu and Siddarth K. Joshi and Mustafa Gundogan and Thomas Brougham and David Lowndes and Luca Mazzarella and Markus Krutzik and Sonali Mohapatra and Daniele Dequal and Giuseppe Vallone and Paolo Villoresi and Alexander Ling and Thomas Jennewein and Makan Mohageg and John Rarity and Ivette Fuentes and Stefano Pirandola and Daniel K. L. Oi},
Title = {Advances in Space Quantum Communications},
Year = {2021},
Eprint = {arXiv:2103.12749},
}

@article{LSU_satsim,
  doi = {10.1038/s41534-020-00327-5},
  %url = {https://doi.org/10.1038/s41534-020-00327-5},
  year = {2021},
  month = jan,
  publisher = {Springer Science and Business Media {LLC}},
  volume = {7},
  number = {1},
  author = {Sumeet Khatri and Anthony J. Brady and Ren{\'{e}}e A. Desporte and Manon P. Bart and Jonathan P. Dowling},
  title = {Spooky action at a global distance: analysis of space-based entanglement distribution for the quantum internet},
  journal = {npj Quantum Information}
}

@article{Pan_exp_crypto,
  doi = {10.1038/s41586-020-2401-y},
  %url = {https://doi.org/10.1038/s41586-020-2401-y},
  year = {2020},
  month = jun,
  publisher = {Springer Science and Business Media {LLC}},
  volume = {582},
  number = {7813},
  pages = {501--505},
  author = {Juan Yin and Yu-Huai Li and Sheng-Kai Liao and Meng Yang and Yuan Cao and Liang Zhang and Ji-Gang Ren and Wen-Qi Cai and Wei-Yue Liu and Shuang-Lin Li and Rong Shu and Yong-Mei Huang and Lei Deng and Li Li and Qiang Zhang and Nai-Le Liu and Yu-Ao Chen and Chao-Yang Lu and Xiang-Bin Wang and Feihu Xu and Jian-Yu Wang and Cheng-Zhi Peng and Artur K. Ekert and Jian-Wei Pan},
  title = {Entanglement-based secure quantum cryptography over 1, 120 kilometres},
  journal = {Nature}
}

@article{Pan_exp_teleport,
  doi = {10.1038/nature23675},
  %url = {https://doi.org/10.1038/nature23675},
  year = {2017},
  month = aug,
  publisher = {Springer Science and Business Media {LLC}},
  volume = {549},
  number = {7670},
  pages = {70--73},
  author = {Ji-Gang Ren and Ping Xu and Hai-Lin Yong and Liang Zhang and Sheng-Kai Liao and Juan Yin and Wei-Yue Liu and Wen-Qi Cai and Meng Yang and Li Li and Kui-Xing Yang and Xuan Han and Yong-Qiang Yao and Ji Li and Hai-Yan Wu and Song Wan and Lei Liu and Ding-Quan Liu and Yao-Wu Kuang and Zhi-Ping He and Peng Shang and Cheng Guo and Ru-Hua Zheng and Kai Tian and Zhen-Cai Zhu and Nai-Le Liu and Chao-Yang Lu and Rong Shu and Yu-Ao Chen and Cheng-Zhi Peng and Jian-Yu Wang and Jian-Wei Pan},
  title = {Ground-to-satellite quantum teleportation},
  journal = {Nature}
}

@article{AntiaQCS,
  doi = {10.1088/1367-2630/11/4/045011},
  %url = {https://doi.org/10.1088/1367-2630/11/4/045011},
  year = {2009},
  month = apr,
  publisher = {{IOP} Publishing},
  volume = {11},
  number = {4},
  pages = {045011},
  author = {Caleb Ho and Ant{\'{\i}}a Lamas-Linares and Christian Kurtsiefer},
  title = {Clock synchronization by remote detection of correlated photon pairs},
  journal = {New Journal of Physics}
}

@article{Lee2019,
  doi = {10.1063/1.5086493},
  %url = {https://aip.scitation.org/doi/10.1063/1.5086493?af=R&},
  year = {2019},
  month = march,
  publisher = {American Institute of Physics ({AIP})},
  volume = {14},
  number = {},
  pages = {101102},
  author = {Jianwei Lee and Lijiong Shen and Alessandro Cerè and James Troupe and Antia Lamas-Linares and Christian Kurtsiefer},
  title = {Symmetrical clock synchronization with time-correlated photon pairs},
  journal = {Applied Physics Letters}
}

@article{Pan_satellite_QCS,
  doi = {10.1038/s41567-020-0892-y},
  %url = {https://doi.org/10.1038/s41567-020-0892-y},
  year = {2020},
  month = may,
  publisher = {Springer Science and Business Media {LLC}},
  volume = {16},
  number = {8},
  pages = {848--852},
  author = {Hui Dai and Qi Shen and Chao-Ze Wang and Shuang-Lin Li and Wei-Yue Liu and Wen-Qi Cai and Sheng-Kai Liao and Ji-Gang Ren and Juan Yin and Yu-Ao Chen and Qiang Zhang and Feihu Xu and Cheng-Zhi Peng and Jian-Wei Pan},
  title = {Towards satellite-based quantum-secure time transfer},
  journal = {Nature Physics}
}

@article{Jon_QCS_2018,
  doi = {10.1038/s41534-018-0090-2},
  %url = {https://doi.org/10.1038/s41534-018-0090-2},
  year = {2018},
  month = aug,
  publisher = {Springer Science and Business Media {LLC}},
  volume = {4},
  number = {1},
  author = {Ebubechukwu O. Ilo-Okeke and Louis Tessler and Jonathan P. Dowling and Tim Byrnes},
  title = {Remote quantum clock synchronization without synchronized clocks},
  journal = {npj Quantum Information}
}

@article{QClock_network_sat,
  doi = {10.1038/nphys3000},
  %url = {https://doi.org/10.1038/nphys3000},
  year = {2014},
  month = jun,
  publisher = {Springer Science and Business Media {LLC}},
  volume = {10},
  number = {8},
  pages = {582--587},
  author = {P. K{\'{o}}m{\'{a}}r and E. M. Kessler and M. Bishof and L. Jiang and A. S. S{\o}rensen and J. Ye and M. D. Lukin},
  title = {A quantum network of clocks},
  journal = {Nature Physics}
}

@article{Qoptics_space_exp,
Author = {David Rideout and Thomas Jennewein and Giovanni Amelino-Camelia and Tommaso F. Demarie and Brendon L. Higgins and Achim Kempf and Adrian Kent and Raymond Laflamme and Xian Ma and Robert B. Mann and Eduardo Martin-Martinez and Nicolas C. Menicucci and John Moffat and Christoph Simon and Rafael Sorkin and Lee Smolin and Daniel R. Terno},
Title = {Fundamental quantum optics experiments conceivable with satellites -- reaching relativistic distances and velocities},
Year = {2012},
Eprint = {arXiv:1206.4949},
Howpublished = {Class. Quantum Grav. 29 224011 (2012)},
Doi = {10.1088/0264-9381/29/22/224011},
}

@article{Einstein_eq_Qopt,
Author = {Daniel R. Terno and Francesco Vedovato and Matteo Schiavon and Alexander R. H. Smith and Piergiovanni Magnani and Giuseppe Vallone and Paolo Villoresi},
Title = {Proposal for an Optical Test of the Einstein Equivalence Principle},
Year = {2018},
Eprint = {arXiv:1811.04835},
}

@article{Bell_test_weak_grav,
  doi = {10.1088/1361-6382/ab8a60},
  %url = {https://doi.org/10.1088/1361-6382/ab8a60},
  year = {2020},
  month = sep,
  publisher = {{IOP} Publishing},
  volume = {37},
  number = {19},
  pages = {195001},
  author = {M Rivera-Tapia and A Delgado and G Rubilar},
  title = {Weak gravitational field effects on large-scale optical interferometric Bell tests},
  journal = {Classical and Quantum Gravity}
}

@article{Anthony_HOM_grav,
  doi = {10.1103/physrevresearch.3.023024},
  %url = {https://doi.org/10.1103/physrevresearch.3.023024},
  year = {2021},
  month = apr,
  publisher = {American Physical Society ({APS})},
  volume = {3},
  number = {2},
  author = {Anthony J. Brady and Stav Haldar},
  title = {Frame dragging and the Hong-Ou-Mandel dip: Gravitational effects in multiphoton interference},
  journal = {Physical Review Research}
}

@article{SAGE,
  doi = {10.1140/epjd/e2019-100324-6},
  %url = {https://doi.org/10.1140/epjd/e2019-100324-6},
  year = {2019},
  month = nov,
  publisher = {Springer Science and Business Media {LLC}},
  volume = {73},
  number = {11},
  author = {Guglielmo M. Tino and Angelo Bassi and Giuseppe Bianco and Kai Bongs and Philippe Bouyer and Luigi Cacciapuoti and Salvatore Capozziello and Xuzong Chen and Maria L. Chiofalo and Andrei Derevianko and Wolfgang Ertmer and Naceur Gaaloul and Patrick Gill and Peter W. Graham and Jason M. Hogan and Luciano Iess and Mark A. Kasevich and Hidetoshi Katori and Carsten Klempt and Xuanhui Lu and Long-Sheng Ma and Holger M\"{u}ller and Nathan R. Newbury and Chris W. Oates and Achim Peters and Nicola Poli and Ernst M. Rasel and Gabriele Rosi and Albert Roura and Christophe Salomon and Stephan Schiller and Wolfgang Schleich and Dennis Schlippert and Florian Schreck and Christian Schubert and Fiodor Sorrentino and Uwe Sterr and Jan W. Thomsen and Giuseppe Vallone and Flavio Vetrano and Paolo Villoresi and Wolf von Klitzing and David Wilkowski and Peter Wolf and Jun Ye and Nan Yu and Mingsheng Zhan},
  title = {{SAGE}: A proposal for a space atomic gravity explorer},
  journal = {The European Physical Journal D}
}

@incollection{Zhu_GPS_relativity,
  doi = {10.1007/bfb0011322},
  %url = {https://doi.org/10.1007/bfb0011322},
  publisher = {Springer-Verlag},
  pages = {41--46},
  author = {S. Y. Zhu and E. Groten},
  title = {Relativistic effects in {GPS}},
  booktitle = {{GPS}-Techniques Applied to Geodesy and Surveying}
}

@article{neutron_gravAbele2012,
  doi = {10.1088/1367-2630/14/5/055010},
  %url = {https://doi.org/10.1088/1367-2630/14/5/055010},
  year = {2012},
  month = may,
  publisher = {{IOP} Publishing},
  volume = {14},
  number = {5},
  pages = {055010},
  author = {Hartmut Abele and Helmut Leeb},
  title = {Gravitation and quantum interference experiments with neutrons},
  journal = {New Journal of Physics}
}

@article{BEC_space_interference_Lachmann2021,
  doi = {10.1038/s41467-021-21628-z},
  %url = {https://doi.org/10.1038/s41467-021-21628-z},
  year = {2021},
  month = feb,
  publisher = {Springer Science and Business Media {LLC}},
  volume = {12},
  number = {1},
  author = {Maike D. Lachmann and Holger Ahlers and Dennis Becker and Aline N. Dinkelaker and Jens Grosse and Ortwin Hellmig and Hauke M\"{u}ntinga and Vladimir Schkolnik and Stephan T. Seidel and Thijs Wendrich and Andr{\'{e}} Wenzlawski and Benjamin Carrick and Naceur Gaaloul and Daniel L\"{u}dtke and Claus Braxmaier and Wolfgang Ertmer and Markus Krutzik and Claus L\"{a}mmerzahl and Achim Peters and Wolfgang P. Schleich and Klaus Sengstock and Andreas Wicht and Patrick Windpassinger and Ernst M. Rasel},
  title = {Ultracold atom interferometry in space},
  journal = {Nature Communications}
}

@article {Xu_sat_grav_decoherence,
	author = {Xu, Ping and Ma, Yiqiu and Ren, Ji-Gang and Yong, Hai-Lin and Ralph, Timothy C. and Liao, Sheng-Kai and Yin, Juan and Liu, Wei-Yue and Cai, Wen-Qi and Han, Xuan and Wu, Hui-Nan and Wang, Wei-Yang and Li, Feng-Zhi and Yang, Meng and Lin, Feng-Li and Li, Li and Liu, Nai-Le and Chen, Yu-Ao and Lu, Chao-Yang and Chen, Yanbei and Fan, Jingyun and Peng, Cheng-Zhi and Pan, Jian-Wei},
	title = {Satellite testing of a gravitationally induced quantum decoherence model},
	volume = {366},
	number = {6461},
	pages = {132--135},
	year = {2019},
	doi = {10.1126/science.aay5820},
	publisher = {American Association for the Advancement of Science},
	issn = {0036-8075},
	%URL = {https://science.sciencemag.org/content/366/6461/132},
	%eprint = {https://science.sciencemag.org/content/366/6461/132.full.pdf},
	journal = {Science}
}

@article{Microscope_Touboul2017,
  doi = {10.1103/physrevlett.119.231101},
  %url = {https://doi.org/10.1103/physrevlett.119.231101},
  year = {2017},
  month = dec,
  publisher = {American Physical Society ({APS})},
  volume = {119},
  number = {23},
  author = {Pierre Touboul and Gilles M{\'{e}}tris and Manuel Rodrigues and Yves Andr{\'{e}} and Quentin Baghi and Joël Berg{\'{e}} and Damien Boulanger and Stefanie Bremer and Patrice Carle and Ratana Chhun and Bruno Christophe and Valerio Cipolla and Thibault Damour and Pascale Danto and Hansjoerg Dittus and Pierre Fayet and Bernard Foulon and Claude Gageant and Pierre-Yves Guidotti and Daniel Hagedorn and Emilie Hardy and Phuong-Anh Huynh and Henri Inchauspe and Patrick Kayser and St{\'{e}}phanie Lala and Claus L\"{a}mmerzahl and Vincent Lebat and Pierre Leseur and Fran{\c{c}}oise Liorzou and Meike List and Frank L\"{o}ffler and Isabelle Panet and Benjamin Pouilloux and Pascal Prieur and Alexandre Rebray and Serge Reynaud and Benny Rievers and Alain Robert and Hanns Selig and Laura Serron and Timothy Sumner and Nicolas Tanguy and Pieter Visser},
  title = {{MICROSCOPE}
		 Mission: First Results of a Space Test of the Equivalence Principle},
  journal = {Physical Review Letters}
}

@article{Vedral_QG_test_PRL,
  title = {Gravitationally Induced Entanglement between Two Massive Particles is Sufficient Evidence of Quantum Effects in Gravity},
  author = {Marletto, C. and Vedral, V.},
  journal = {Phys. Rev. Lett.},
  volume = {119},
  issue = {24},
  pages = {240402},
  numpages = {5},
  year = {2017},
  month = {Dec},
  publisher = {American Physical Society},
  doi = {10.1103/PhysRevLett.119.240402},
  %url = {https://link.aps.org/doi/10.1103/PhysRevLett.119.240402}
}

@article{SBose_QG_test_PRL,
  title = {Spin Entanglement Witness for Quantum Gravity},
  author = {Bose, Sougato and Mazumdar, Anupam and Morley, Gavin W. and Ulbricht, Hendrik and Toro\ifmmode \check{s}\else \v{s}\fi{}, Marko and Paternostro, Mauro and Geraci, Andrew A. and Barker, Peter F. and Kim, M. S. and Milburn, Gerard},
  journal = {Phys. Rev. Lett.},
  volume = {119},
  issue = {24},
  pages = {240401},
  numpages = {6},
  year = {2017},
  month = {Dec},
  publisher = {American Physical Society},
  doi = {10.1103/PhysRevLett.119.240401},
  %url = {https://link.aps.org/doi/10.1103/PhysRevLett.119.240401}
}

@article{Newbury2016_freq_combs,
  doi = {10.1103/physrevx.6.021016},
  %url = {https://doi.org/10.1103/physrevx.6.021016},
  year = {2016},
  month = may,
  publisher = {American Physical Society ({APS})},
  volume = {6},
  number = {2},
  author = {Jean-Daniel Desch{\^{e}}nes and Laura C. Sinclair and Fabrizio R. Giorgetta and William C. Swann and Esther Baumann and Hugo Bergeron and Michael Cermak and Ian Coddington and Nathan R. Newbury},
  title = {Synchronization of Distant Optical Clocks at the Femtosecond Level},
  journal = {Physical Review X}
}

@article{COW,
  author={Colella, R. and Overhauser, A.W. and Werner, S.A.},
  journal={Phys. Rev. Lett.},
  title={Observation of Gravitationally Induced Quantum Interference}, 
  year={1975},
  volume={34},
  pages={1472},
  doi={10.1103/PhysRevLett.34.1472}}

@article{Taylor2020,
  author={Taylor, M.T. and Belmonte, A. and Holberg, L. and Kahn, J.},
  journal={Phys. Rev. A},
  title={Effect of atmospheric turbulence on timing instability for partially reciprocal two-way optical time transfer links}, 
  year={2020},
  volume={101},
  pages={033843},
  doi={10.1103/PhysRevA.101.033843}}

@article{Belmonte2017,
  author={Belmonte, A. and Taylor, M.T. and Holberg, L. and Kahn, J.},
  journal={Optics Express},
  title={Effect of atmospheric anisoplanatism on earth-to-satellite time transfer over laser communication links}, 
  year={2017},
  volume={25},
  pages={15676},
  doi={10.1364/OE.25.015676}}

@Article{Shih2004,
  author   = {Alejandra Valencia and Giuliano Scarcelli and Yanhua Shih},
  title    = {Distant clock synchronization using entangled photon pairs},
  journal  = {Applied Physics Letters},
  year     = {2004},
  volume   = {85},
  pages    = {2655},
  doi      = {10.1063/1.1797561},
}

@article{spdc_time_corr1970,
  doi = {10.1103/physrevlett.25.84},
  %url = {https://doi.org/10.1103/physrevlett.25.84},
  year = {1970},
  month = jul,
  publisher = {American Physical Society ({APS})},
  volume = {25},
  number = {2},
  pages = {84--87},
  author = {David C. Burnham and Donald L. Weinberg},
  title = {Observation of Simultaneity in Parametric Production of Optical Photon Pairs},
  journal = {Physical Review Letters}
}

@article{DSAC2021,
  doi = {10.1038/s41586-021-03571-7},
  %url = {https://www.nature.com/articles/s41586-021-03571-7},
  year = {2021},
  month = jun,
  publisher = {},
  volume = {595},
  number = {},
  pages = {43--47},
  author = {E.A. Burt and J.D. Prestage and R.L. Tjoelker and D.G. Enzer and D. Kuang and D.W. Murphy and D.E. Robison and J.M. Seubert and R.T. Wang and T.A. Ely},
  title = {Demonstration of a trapped-ion atomic clock in space},
  journal = {Nature}
}

@article{DARPA2019,
  doi = {},
  %url = {https://www.darpa.mil/news-events/2019-08-20},
  year = {2019},
  month = aug,
  publisher = {DARPA},
  volume = {},
  number = {},
  pages = {},
  author = {},
  title = {DARPA Making Progress on Miniaturized Atomic Clocks for Future PNT Applications},
  journal = {DARPA Website}
}

@article{GPSWorld2011,
  doi = {},
  url = {https://www.gpsworld.com/defense-warfighter-microtechnology-comes-age/},
  year = {2011},
  month = sep,
  publisher = {},
  volume = {},
  number = {},
  pages = {},
  author = {Andrei M. Shkel (DARPA)},
  title = {Microtechnology Comes of Age},
  journal = {GPS World}
}

@article{NIST2020,
  doi = {},
  url = {https://www.gpsworld.com/defense-warfighter-microtechnology-comes-age/},
  year = {2020},
  month = dec,
  publisher = {},
  volume = {},
  number = {},
  pages = {},
  author = {Laura Ost},
  title = {Success Story: Chip-Scale Atomic Clock},
  journal = {NIST Webpage}
}

@article{GE_2021,
  doi = {10.1116/5.0051881},
  url = {},
  year = {2021},
  month = jul,
  publisher = {AIP Publishing},
  volume = {3},
  number = {},
  pages = {030501},
  author = {Stephen Bush and William Challener and Guillaume Mantelet},
  title = {A perspective on industrial quantum networks},
  journal = {AVS Quantum Science}
}

@article{Dolinar2011,
  doi = {10.1109/ICSOS.2011.5783682},
  url = {},
  year = {2011},
  month = jun,
  publisher = {IEEE},
  volume = {},
  number = {},
  pages = {269-278},
  author = {Sam Dolinar and Kevin M. Birnbaum and Baris I. Erkmen and Bruce Moision},
  title = {On approaching the ultimate limits of photon-efficient and bandwidth-efficient optical communication},
  journal = {2011 International Conference on Space Optical Systems and Applications (ICSOS)}
}

@misc{wiki:Network_Time_Protocol,
   author = "Wikipedia",
   title = "{Network Time Protocol} --- {W}ikipedia{,} The Free Encyclopedia",
   year = "2021",
   howpublished = {\url{http://en.wikipedia.org/w/index.php?title=Network\%20Time\%20Protocol&oldid=1048423448}},
   note = "[Online; accessed 10-October-2021]"
 }

@ARTICLE{NTP_Mills,
  author={Mills, D.L.},
  journal={IEEE Transactions on Communications}, 
  title={Internet time synchronization: the network time protocol}, 
  year={1991},
  volume={39},
  number={10},
  pages={1482-1493},
  doi={10.1109/26.103043}}

@misc{NTP_security,
	series =	{Request for Comments},
	number =	5906,
	howpublished =	{RFC 5906},
	publisher =	{RFC Editor},
	doi =		{10.17487/RFC5906},
	url =		{https://rfc-editor.org/rfc/rfc5906.txt},
        author =	{Professor David L. Mills and Brian Haberman},
	title =		{{Network Time Protocol Version 4: Autokey Specification}},
	pagetotal =	58,
	year =		2010,
	month =		jun,
}

@article{NTP_Attacks,
  title={Attacking the Network Time Protocol},
  journal={IACR Cryptology ePrint Archive},
  volume={2015},
  pages={1020},
  url={https://eprint.iacr.org/2015/1020},
  author={Aanchal Malhotra and Isaac E. Cohen and Erik Brakke and Sharon Goldberg},
  year=2015
}

@article{Bedington2016,
  doi = {10.1140/epjqt/s40507-016-0051-7},
  url = {https://epjquantumtechnology.springeropen.com/articles/10.1140/epjqt/s40507-016-0051-7},
  year = {2016},
  month = oct,
  publisher = {American Institute of Aeronautics and Astronautics ({AIAA})},
  volume = {3},
  number = {12},
  pages = {},
  author = {Robert Bedington and Xueliang Bai and Edward Truong-Cao and Yue Chuan Tan and Kadir Durak and Aitor Villar Zafra and James A Grieve and Daniel KL Oi and Alexander Ling},
  title = {Nanosatellite experiments to enable future space-based QKD missions},
  journal = {EPJ Quantum Technology}
}

@article{Kimble_quantum_internet,
  doi = {10.1038/nature07127},
  url = {https://doi.org/10.1038/nature07127},
  year = {2008},
  month = jun,
  publisher = {Springer Science and Business Media {LLC}},
  volume = {453},
  number = {7198},
  pages = {1023--1030},
  author = {H. J. Kimble},
  title = {The quantum internet}
}

@article{Ashby_GPS_relativity,
  doi = {10.12942/lrr-2003-1},
  url = {https://doi.org/10.12942/lrr-2003-1},
  year = {2003},
  month = jan,
  publisher = {Springer Science and Business Media {LLC}},
  volume = {6},
  number = {1},
  author = {Neil Ashby},
  title = {Relativity in the Global Positioning System}
}
\end{document}